\documentclass[runningheads]{llncs}
\usepackage{graphicx}
\usepackage{tabularx}
\usepackage{multirow}
\usepackage{amsmath}
\usepackage{amsfonts}
\newcommand{\myparagraph}[1]{\smallskip\noindent\textbf{#1}}
\newcounter{daggerfootnote}
\newcommand*{\daggerfootnote}[1]{%
    \setcounter{daggerfootnote}{\value{footnote}}%
    \renewcommand*{\thefootnote}{\fnsymbol{footnote}}%
    \footnote[4]{#1}%
    \setcounter{footnote}{\value{daggerfootnote}}%
    \renewcommand*{\thefootnote}{\arabic{1}}%
    }
\begin{document}
\title{Attention-based Multi-scale Gated Recurrent Encoder with Novel Correlation Loss for COVID-19 Progression Prediction}

\titlerunning{Multi-scale GRU Encoder framework with Correlation Loss}

\author{Aishik Konwer\inst{1}, Joseph Bae\inst{2}, Gagandeep Singh\inst{3}, Rishabh Gattu\inst{3}, Syed Ali\inst{3}, Jeremy Green\inst{3}, Tej Phatak\inst{3}, Prateek Prasanna\inst{2}}


\authorrunning{A. Konwer et al.}

\institute{Department of Computer Science, Stony Brook University, NY, USA
\and Department of Biomedical Informatics, Stony Brook University, NY, USA
\and Department of Radiology, Newark Beth Israel Medical Center, NJ, USA
\email{akonwer@cs.stonybrook.edu, prateek.prasanna@stonybrook.edu}}

\maketitle{}              
\begin{abstract}
COVID-19 image analysis has mostly focused on diagnostic tasks using single timepoint scans acquired upon disease presentation or admission. We present a deep learning-based approach \protect\daggerfootnote{This work will be presented at MICCAI 2021} to predict lung infiltrate progression from serial chest radiographs (CXRs) of COVID-19 patients. Our method first utilizes convolutional neural networks (CNNs) for feature extraction from patches within the concerned lung zone, and also from neighboring and remote boundary regions. The framework further incorporates a multi-scale Gated Recurrent Unit (GRU) with a correlation module for effective predictions. The GRU accepts CNN feature vectors from three different areas as input and generates a fused representation. The correlation module attempts to minimize the correlation loss between hidden representations of concerned and neighboring area feature vectors, while maximizing the loss between the same from concerned and remote regions. Further, we employ an attention module over the output hidden states of each encoder timepoint to generate a context vector. This vector is used as an input to a decoder module to predict patch severity grades at a future timepoint. Finally, we ensemble the patch classification scores to calculate patient-wise grades. Specifically, our framework predicts zone-wise disease severity for a patient on a given day by learning representations from the previous temporal CXRs. Our novel multi-institutional dataset comprises sequential CXR scans from N=93 patients. Our approach outperforms transfer learning and radiomic feature-based baseline approaches on this dataset.

\keywords{COVID-19 \and Correlation \and Attention \and Gated Recurrent Unit \and Transfer Learning.}
\end{abstract}

\section{Introduction}
Coronavirus disease 2019 (COVID-19) remains at the forefront of threats to public health. As a result, there continues to be a critical need to further understand the progression of the disease process. In the United States, chest radiographs (CXRs) are the most commonly used imaging modality for the monitoring of COVID-19. On CXR, COVID-19 infection has been found to manifest as opacities within lung regions. Previous studies have demonstrated that the location, extent, and temporal evolution of these findings can be correlated to disease progression \cite{toussie_clinical_2020}. Studies have shown that COVID-19 infection frequently results in bilateral lower lung opacities on CXR and that these opacities may migrate to other lung regions throughout the disease's clinical course. \cite{toussie_clinical_2020,litmanovich_review_2020}. This suggests that COVID-19 progression may be appreciable on CXR via examination of the spatial spread of radiographic findings across multiple timepoints. 

Despite the many studies analyzing the use of CXRs in COVID-19, machine learning applications have been limited to diagnostic tasks including differentiating COVID-19 from viral pneumonia and predicting clinical outcomes such as mortality and mechanical ventilation requirement \cite{kwon_combining_2020,hu_role_2021}. Many of these studies have reported high sensitivities and specificities for the studied outcomes, but they remain constrained due to deficiencies in publicly available datasets \cite{lopez-cabrera_current_2021}. Furthermore, none have attempted to computationally model the temporal progression of COVID-19 from an imaging perspective. Significantly, most studies have also not explicitly taken into account the spatial evolution of CXR imaging patterns within lung regions that have been demonstrated to correlate with disease severity and progression \cite{toussie_clinical_2020,litmanovich_review_2020}. In this study we take advantage of a unique longitudinal COVID-19 CXR dataset and propose a novel deep learning (DL) approach that exploits the spatial and temporal dependencies of CXR findings in COVID-19 to predict disease progression.    

\begin{figure}
\includegraphics[width=60mm]{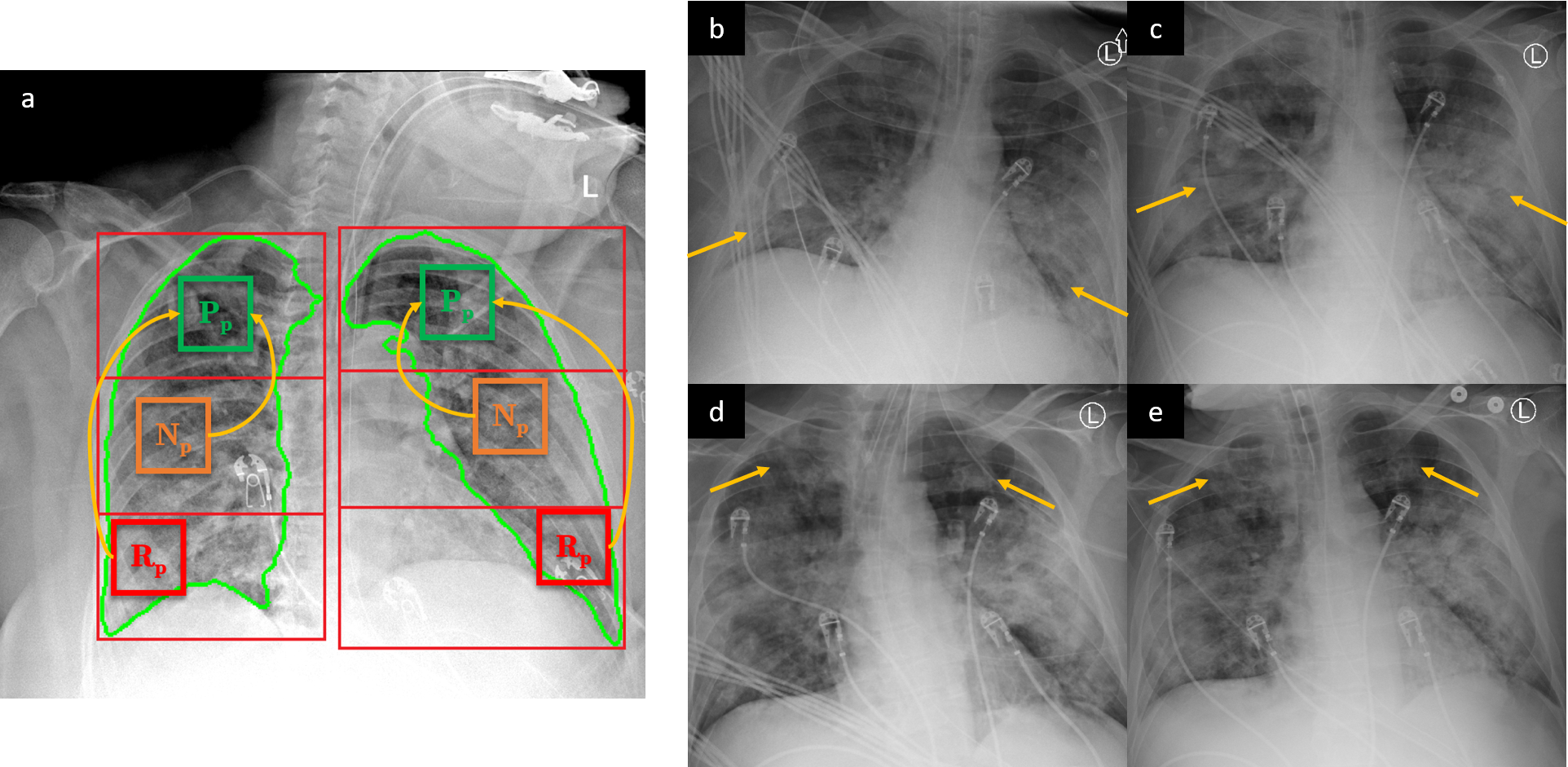}
\centering
\caption{(a) depicts a CXR in which lung fields have been divided into three equal zones. Disease information in patches from primary zone ($Pp$) are more similar to those from neighboring zone ($Np$) than the remote zones ($Rp$). (b)-(e) depict serial CXRs taken for one patient over several days of COVID-19 infection. We note a progression of imaging findings beginning with lower lobe involvement in (b) with spread to middle lung involvement in (c) and upper lung region involvement in (d) and (e).} \label{fig1}
\end{figure}

Previous deep learning (DL) based COVID-19 studies have mainly considered single timepoint CXRs \cite{bae_predicting_2020,shi_review_2020}. Unlike these studies, we analyze CXRs from multiple timepoints to capture lung infiltrate progression. Recurrent neural networks (RNNs) have been widely employed for time series prediction tasks in computer vision problems. Recently, RNNs have also found success in analyzing tumor evolution \cite{tumor} and treatment response from serial medical images~\cite{serial1,serial2}. A Gated Recurrent Unit (GRU) is an RNN which controls information flow using two gates - a Reset gate and an Update gate. Thus, relevant information from past timepoints are forwarded to future timepoints in the form of hidden states. GRUs have been used extensively to predict disease progression \cite{Disease_GRU}. 

In this work, we aim to explore how the different zones of an image are correlated to each other. Many studies have demonstrated the spatial progression of COVID-19 seen on CXR imaging with lung opacities generally being noted in lower lung regions in earlier disease stages before gradually spreading to involve other areas such as the middle and upper lung \cite{toussie_clinical_2020,wong_frequency_2020,litmanovich_review_2020}. Therefore, two neighboring lung zones should have a higher similarity measure than two far-apart zones. Unlike previous approaches, we propose a multi-scale GRU \cite{multimodal} which can accept three distinct inputs at the same timepoint. Apart from primary patches $Pp$ of concerned zone, patches from Neighbor $Np$ and Remote areas $Rp$ are also used as inputs to a GRU cell at a certain timepoint. We include a Correlation module to maximize the correlation measure between $Pp$ and $Np$, while minimizing the correlation between $Pp$ and $Rp$. Finally, an attention layer is applied over hidden states to obtain patch weights and give relative importance to patches collected from multiple timepoints. The major contributions of this paper are the following: (1) Our work uses a multi-scale GRU framework to model the progression of lung infiltrates over multiple timepoints to predict the severity of imaging infiltrates at a later stage. (2) Disease patterns in adjacent regions tend to be spatially related to each other. COVID-19 imaging infiltrates exhibit similar patterns of correlation across lung regions on CXRs. We are the first to use a dedicated correlation module within our temporal encoder that exploits this latent state inter-zone similarity with a novel correlation loss.

\section{Methodology}
Varying numbers of temporal images are available for each patient. The number of timepoints is equal to $d$ which may vary from 4 to 13 for a given patient. The images corresponding to these $d$ timepoints are denoted by $I_{t_1},I_{t_2},...I_{t_{d-1}},I_{t_d}$. The left and right lung masks are generated from these images using a residual U-Net model \cite{bae_predicting_2020}. These masks are each further subdivided into 3 lung zones - Upper ($L_1, R_1$), Middle ($L_2, R_2$) and Lower ($L_3, R_3$) zones. Our collaborating radiologists assigned severity grades to each of the 6 zones as $g_0=0$, $g_1=1$ or $g_2=2$ depending on the zonal infiltrate severity. This procedure mirrors the formulation of other scoring systems \cite{kwon_combining_2020}.
We train $6$ different models for each of the six zones - $M_{L_1}$, $M_{L_2}$, $M_{L_3}$, $M_{R_1}$, $M_{R_2}$, and $M_{R_3}$. We adopt this zone-wise granular approach to overcome the need of image registration. 

\subsection{Overview}
We implement an Encoder-Decoder framework based on seq2seq model \cite{Sequence} in order to learn sequence representations. Specifically, our framework includes two recurrent neural networks: a multi-scale encoder and a decoder. The training of the multi-scale encoder involves fusion of three input patches each from $Pp$, $Np$ and $Rp$ - concerned (current zone of interest), neighboring, and remote zones, at each of the timepoints, to generate a joint feature vector. The attention weighted context vector that we obtain from the encoder is finally used as input to the decoder. The decoder at its first timepoint attempts to classify this encoder context vector into the 3 severity labels. The multi-scale encoder is trained with the help of a correlation module to retain only relevant information from each of the patches of three distinct zones.

\subsection{Patch extraction}
Each image zone is divided into sixteen square grids. These grids are resized to dimension $128\times128$ and used as primary patches $Pp$ for the concerned zone. Now, for each zone, we also consider 8 patches from the boundary of two adjoining neighbor zones. For example, in the case of $L_1$ zone, we use 4 patch grids from $R_1$ boundary and 4 patch grids from $L_2$ boundary. Similarly, in the case of middle zone $L_2$, we use 4 patch grids each from nearest $L_1$ and $L_3$ boundaries. Thus we build a pool of 8 neighboring $Np$ patches for each concerned lung zone. Additionally, we also create a cluster of 8 $Rp$ patches coming from the far-away boundaries of remote zones. e.g. $Rp$ patches for $L_1$ is collected from boundaries of $L_3$.
\subsection{Feature extraction}
For a particular model, say $M_{L_1}$, each $Pp$ patch from zone $L_1$ is fed as input to a Convolutional neural network (CNN) to predict the severity scores at a given timepoint. Similarly, one random patch from each of $Np$ and $Rp$, are also passed into the same CNN. As an output of the CNN, we obtain three $1\times256$ dimensional feature vectors. The CNN network configuration contains five convolutional layers, each associated with an operation of max-pooling. The network terminates with a fully connected layer.

\subsection{Encoder}
\myparagraph{Multi-scale GRU.}
The GRU module used here is a multi-scale extension of the standard GRU. It houses different gating units - the reset gate and the update gate which control the flow of relevant information in a GRU. The GRU module takes $P_t$, $N_t$, and $R_t$ as inputs (denoted by $X^i_t$, where i=1,2,3) at time step $t$ and monitors four latent variables, namely the joint representation $h_t$, and input-specific representations $h^1_t$, $h^2_t$ and $h^3_t$. The fused representation $h_t$ is actually treated as a single descriptor for the multi-input data that helps in learning the temporal context of our data over multiple timepoints. The input-specific representations $h^1_t$, $h^2_t$ and $h^3_t$ constitute the projections of three distinct inputs. They are used to calculate two correlation measures among them in the GRU module. The computation within this module may be formally expressed as follows:

\begin{equation}
r^i_t = \sigma(W^i_r X^i_t + U_r h_{t-1} + b^i_r) 
\end{equation}
\begin{equation}
z^i_t = \sigma(W^i_z X^i_t + U_z h_{t-1} + b^i_z) 
\end{equation}
\begin{equation}
\tilde{h}^i_t = \varphi(W^i_h X^i_t + U_h (r^i_t\odot h_{t-1}) + b^i_h) , i = 1,2,3
\end{equation}

\begin{equation}
r_t = \sigma(\sum_{i=1}^{3}w^i_t(W^i_r X^i_t + b^i_r) + U_r h_{t-1}) \end{equation}
\begin{equation}
z_t = \sigma(\sum_{i=1}^{3}w^i_t(W^i_z X^i_t + b^i_z) + U_z h_{t-1}) \end{equation}
\begin{equation}
\tilde{h}_t = \varphi(\sum_{i=1}^{3}w^i_t(W^i_h X^i_h + b^i_h) + U_h (r_t\odot h_{t-1} ) 
\end{equation}

\begin{equation}
h^i_t = (1 - z^i_t) \odot h_{t-1} + z^i_t \odot \tilde{h}^i_t, 
i = 1,2,3
\end{equation}
\begin{equation}
h_t = (1 - z_t) \odot h_{t-1} + z_t \odot \tilde{h}_t
\end{equation}

where $\sigma$ is the logistic sigmoid function and $\varphi$ is the hyperbolic tangent function, $r$ and $z$ are the input to the reset and update gates, and $h$ and $\tilde{h}$ represent the activation and candidate activation, respectively, of the standard GRU \cite{GRU}. $W_r$, $W_z$, $W_h$, $U_r$, $U_z$ and $U_h$ are the weight parameters learned during training. $w^i_t$ (i=1,2,3) are also learned parameters. $b_r$, $b_z$ and $b_h$ are the biases. $X^i_t$ (i=1,2,3) are the CNN feature vectors of patches from the three zones - $P_p$, $N_p$ and $R_p$.


\myparagraph{Correlation module.}
In order to obtain a better joint representation for temporal learning, we introduce an important component into the multi-scale encoder, one that explicitly captures the correlation between the three distinct inputs. Our model explicitly applies a correlation-based loss term in the fusion process. The principle of our model is to maximize the correlation between features from $Pp$ and $Np$, and to minimize the correlation between features from $Pp$ and $Rp$. Pearson coefficient has been used to compute the correlation. Hence this module computes the correlation between the projections $h^1_t$, $h^2_t$ and also between $h^1_t$, $h^3_t$ obtained from the GRU module. We denote the correlation-based loss function as 
\begin{equation}
L_{corr} = max[corr(h^1_t,h^2_t)] + min[corr(h^1_t,h^3_t)]
\end{equation}

For all patients, independently for each patch from $Pp$ and $Np$ zones, we maximized the correlation function. Similarly we minimized the correlation function for each patch from $Pp$ and $Rp$ zones.

\myparagraph{Attention module.}
The hidden state from each GRU cell is passed through an attention network. The attention weights $\alpha_1$, $\alpha_2$,...,$\alpha_{d-1}$ are computed for each timepoint. These scores are then fed to a softmax layer to obtain the probability weight distribution,
such that the summation of all attention weights covering the available $d-1$ timepoints of the encoder equals to 1. We compute a weighted summation of these attention weights and the GRU hidden states' vectors to construct a holistic context vector for the encoder output.

\subsection{Decoder}
The attention weighted context vector obtained from $d-1$ timepoints of the encoder is used as an input to the decoder. A linear classifier and softmax layer is applied on the GRU decoder's hidden state to obtain three severity scores - $g_0$, $g_1$, and $g_2$. For each patient and zone, we predict 16 such patch classification scores for the ${I_{t_d}}^{th}$ image. We employ majority voting as an ensemble procedure on these scores to obtain the final patient-wise grade.

\begin{figure}
\centering
\includegraphics[width=95mm]{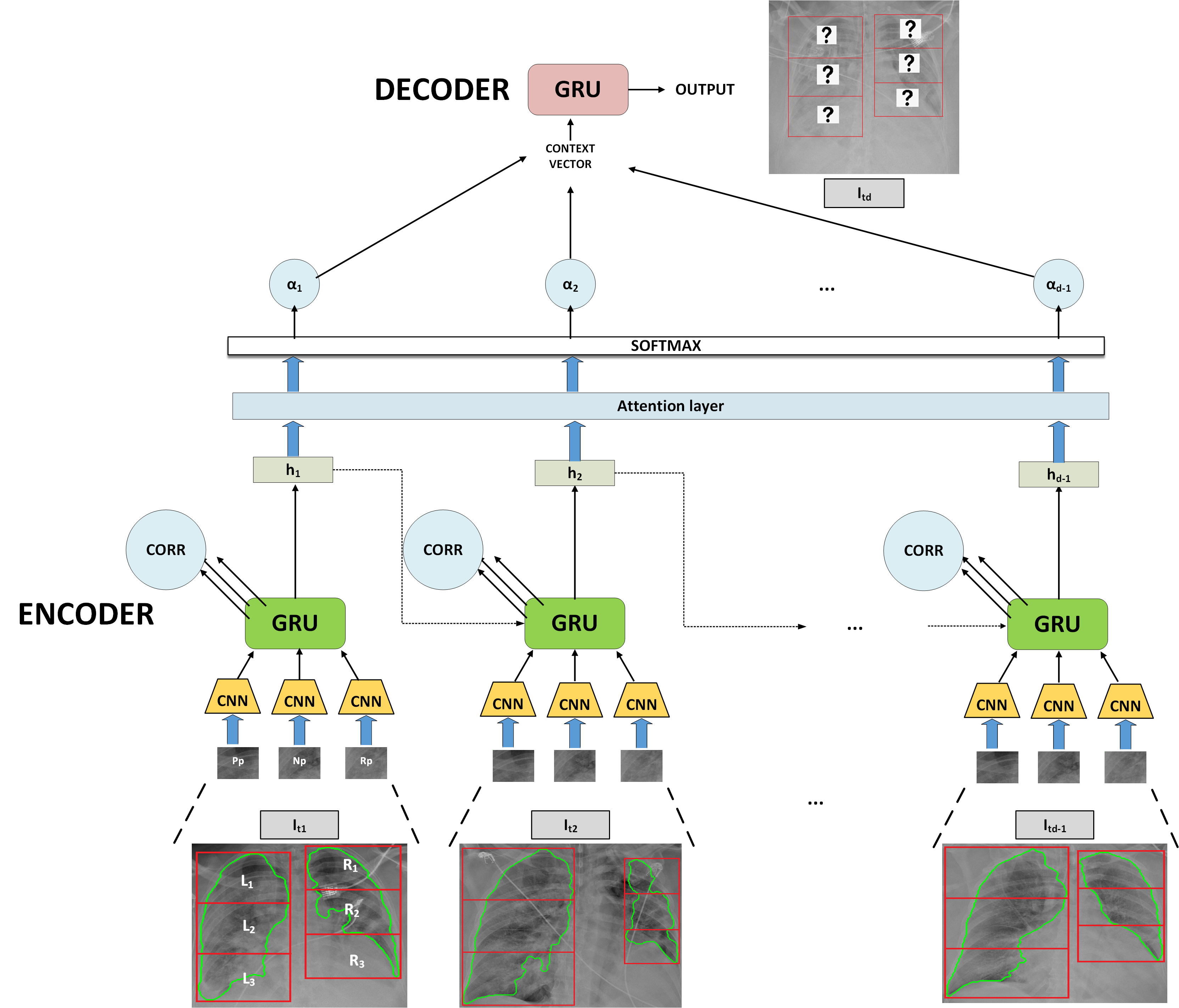}
\caption{\textbf{Architecture of the proposed approach}. We show here model $M_{L_1}$ which deals with patches from $L_1$ zone. At each timepoint, 3 patches each from $Pp$, $Np$ and $Rp$ are inputs to CNN network. The generated CNN features are passed into a GRU cell. Fused hidden state GRU output $h_t$ is used to calculate attention weights. Attention weighted summation of multiple such hidden states form the context vector for decoding purpose.} \label{fig2}
\end{figure}

\section{Experimental Design}
\subsection{Dataset Description}
Our multi-institutional dataset, COVIDProg \cite{midl}, contains 621 antero-posterior CXR scans from 93 COVID-19 patients, collected from multiple days. 23 cases were obtained from \textit{Newark Beth Israel Medical Center}. The remaining 70 cases curated from \textit{Stony Brook University Hospital}. All the CXRs were of dimension $3470\times4234$. Additional details can be found in Supplementary section 3.

\subsection{Implementation Details}
For training the CNN and GRU, a cross entropy loss function was used along with the designed correlation loss discussed earlier. Optimization of the network was done using Adam Optimizer. Each of the 6 models is trained once for 300 iterations with a batch size of 30 and a learning rate of 0.001. The total number of epochs is 20. We used pack padded sequence to mask out all losses that surpassed the required sequence length. Thus, we could nullify the effect of missing timesteps for a patient in the dataset. We have adopted a 5-fold cross validation approach to predict the $I_{t_d}$ th image severity grades for 93 patients, using $d-1$ images as encoder input.

\myparagraph{First baseline approach ($B_1$).}
We trained 6 different models based on a transfer learning based framework, illustrated in Supplementary section 4. All the pretrained convolutional weights of a VGG-16 network \cite{vgg16}  were kept same. The last two layers of the network were replaced with two new fully connected layers to deal with the 3-class classification problem. For a particular model, $M_{L_1}$, $64\times64$ dimension patches were extracted from the $L_1$ zone using a sliding window approach with a stride size 32. After passing these patches as input to our VGG-16, we obtained a $P\times4096$ feature vector where $P$ denotes the total number of patches extracted for a patient from the $L_1$ zones of images collected from multiple timepoints $t_1$, $t_2$,...,$t_{d-1}$. We used a simple feature averaging technique to obtain a $1\times4096$ feature vector from the $P\times4096$ feature for each patient. Finally a 1-D neural network was trained to classify the patches into severity grades $g_0$, $g_1$ and $g_2$ predictions for $I_{t{d}}$th image. Majority voting was used as an ensemble procedure to convert these patch classification grades to a patient-wise grade. 

\myparagraph{Second baseline approach ($B_2$).}
We built a radiomic feature based pipeline. 445 texture-based radiomic features \cite{pyrad} were extracted from the concerned lung zone. These features were similarly averaged into a single feature vector and classified using random forest classifier.
\section{Results}
 Averaged results are presented after 5 runs of model-testing. Accuracy is computed for each of the 6 lung zones, while the precision and recall are measured for each of the severity grades, $g_0$, $g_1$ and $g_2$. The results using our approach and the two baseline methods are illustrated in Tables \ref{Left results} and \ref{Right results}  for the left and the right lung zones, respectively. In all the zones, except $R_3$, our method performed significantly better than both baseline approaches. For example, in left lung upper zone, we achieved an accuracy of 75.26\%. The baseline accuracies were 60.21\% and 56.98\% for $B_1$ and $B_2$, respectively.

\begin{table*}[!t]
\caption{Quantitative Results on Left lung zones}
\label{Left results}
\resizebox{\textwidth}{!}{
\begin{tabular}{|c|c|c|c|c|c|c|c|c|c|c|c|c|c|c|c|c|c|c|c|c|c|}
\hline
\multirow{2}{*}{\textbf{Methods}} & \multicolumn{7}{c|}{\textbf{Left Lung Upper}} & \multicolumn{7}{c|}{\textbf{Left Lung Middle}} & \multicolumn{7}{c|}{\textbf{Left Lung Lower}} \\ \cline{2-22} 
 & \textbf{$Acc (\%)$} & \multicolumn{3}{c|}{\textbf{\begin{tabular}[c]{@{}c@{}}$Pre$\\ 0    1    2\end{tabular}}} & \multicolumn{3}{c|}{\textbf{\begin{tabular}[c]{@{}c@{}}$Rec$\\ 0    1    2\end{tabular}}} & \textbf{$Acc (\%)$} & \multicolumn{3}{c|}{\textbf{\begin{tabular}[c]{@{}c@{}}$Pre$\\ 0    1    2\end{tabular}}} & \multicolumn{3}{c|}{\textbf{\begin{tabular}[c]{@{}c@{}}$Rec$\\ 0    1    2\end{tabular}}} & \textbf{$Acc (\%)$} & \multicolumn{3}{c|}{\textbf{\begin{tabular}[c]{@{}c@{}}$Pre$\\ 0    1    2\end{tabular}}} & \multicolumn{3}{c|}{\textbf{\begin{tabular}[c]{@{}c@{}}$Rec$\\ 0    1    2\end{tabular}}} \\ \hline
Baseline-1 & 60.21 & 0.53 & 0.72 & 0.53 & 0.43 & 0.51 & \textbf{0.77} & 62.36 & 0.55 & 0.69 & 0.71 & \textbf{0.65} & 0.63 & 0.60 & 59.13 & 0.53 & 0.67 & 0.72 & 0.56 & 0.69 & 0.80 \\ \hline
Baseline-2 & 56.98 & 0.54 & 0.60 & 0.52 & 0.63 & 0.56 & 0.66 & 54.83 & 0.58 & 0.74 & 0.67 & 0.50 & 0.71 & 0.61 & 60.21 & \textbf{0.74} & 0.69 & 0.76 & 0.45 & 0.69 & 0.73 \\ \hline
\begin{tabular}[c]{@{}c@{}}Variant-1 \end{tabular} & 66.66 & 0.54 & 0.68 & \textbf{0.72} & 0.67 & 0.73 & 0.56 & 67.74 & 0.61 & 0.69 & 0.63 & 0.54 & 0.69 & 0.64 & 63.44 & 0.65 & 0.72 & 0.79 & 0.67 & 0.72 & 0.75\\ \hline
Variant-2 & 73.11 & 0.59 & 0.71 & 0.66 & 0.66 & 0.77 & 0.69 & 70.96 & 0.63 & 0.74 & 0.69 & 0.56 & 0.67 & 0.59 & 68.89 & 0.61 & 0.76 & 0.81 & \textbf{0.70} & \textbf{0.74} & 0.67\\ \hline
Our Approach & \textbf{75.26} & \textbf{0.69} & \textbf{0.73} & 0.58 & \textbf{0.68} & \textbf{0.81} & 0.75 & \textbf{72.04} & \textbf{0.72} & \textbf{0.82} & \textbf{0.77} & 0.63 & \textbf{0.84} & \textbf{0.65} & \textbf{73.11} & 0.66 & \textbf{0.78} & \textbf{0.83} & 0.69 & 0.72 & \textbf{0.85} \\ \hline
\end{tabular}}
\end{table*}

\begin{table*}[!t]
\caption{Quantitative Results on Right lung zones}
\label{Right results}
\resizebox{\textwidth}{!}{\begin{tabular}{|c|c|c|c|c|c|c|c|c|c|c|c|c|c|c|c|c|c|c|c|c|c|}
\hline
\multirow{2}{*}{\textbf{Methods}} & \multicolumn{7}{c|}{\textbf{Right Lung Upper}} & \multicolumn{7}{c|}{\textbf{Right Lung Middle}} & \multicolumn{7}{c|}{\textbf{Right Lung Lower}} \\ \cline{2-22} 
 & \textbf{$Acc (\%)$} & \multicolumn{3}{c|}{\textbf{\begin{tabular}[c]{@{}c@{}}$Pre$\\ 0    1    2\end{tabular}}} & \multicolumn{3}{c|}{\textbf{\begin{tabular}[c]{@{}c@{}}$Rec$\\ 0    1    2\end{tabular}}} & \textbf{$Acc (\%)$} & \multicolumn{3}{c|}{\textbf{\begin{tabular}[c]{@{}c@{}}$Pre$\\ 0    1    2\end{tabular}}} & \multicolumn{3}{c|}{\textbf{\begin{tabular}[c]{@{}c@{}}$Rec$\\ 0    1    2\end{tabular}}} & \textbf{$Acc (\%)$} & \multicolumn{3}{c|}{\textbf{\begin{tabular}[c]{@{}c@{}}$Pre$\\ 0    1    2\end{tabular}}} & \multicolumn{3}{c|}{\textbf{\begin{tabular}[c]{@{}c@{}}$Rec$\\ 0    1    2\end{tabular}}} \\ \hline
Baseline-1 & 61.29 & 0.54 & 0.62 & 0.54 & 0.59 & 0.64 & 0.63 & 62.36 & 0.60 & 0.63 & \textbf{0.67} & 0.61 & 0.64 & 0.59 & 58.06 & 0.71 & 0.58 & 0.64 & 0.52 & 0.63 & 0.73 \\ \hline
Baseline-2 & 55.91 & 0.50 & 0.66 & 0.57 & 0.61 & 0.58 & 0.66 & 59.13 & 0.63 & 0.60 & 0.65 & 0.64 & 0.55 & 0.56 & 53.76 & 0.74 & 0.51 & 0.58 & 0.57 & 0.66 & 0.70 \\ \hline
\begin{tabular}[c]{@{}c@{}}Variant-1 \end{tabular} & 69.89 & 0.55 & 0.64 & \textbf{0.65} & 0.60 & \textbf{0.73} & 0.71 & \textbf{67.74} & 0.68 & 0.63 & 0.58 & 0.68 & 0.74 & 0.68 & 63.44 & 0.71 & 0.63 & 0.57 & \textbf{0.58} & 0.73 & 0.77\\ \hline
Variant-2 & 73.11 & 0.62 & 0.69 & 0.64 & 0.63 & 0.71 & 0.75 & 64.51 & 0.67 & \textbf{0.66} & 0.61 & 0.65 & \textbf{0.76} & 0.70 & 65.59 & 0.73 & 0.68 & \textbf{0.77} & 0.54 & 0.72 & 0.80 \\ \hline
Our Approach & \textbf{76.34} & \textbf{0.67} & \textbf{0.72} & 0.56 & \textbf{0.71} & 0.70 & \textbf{0.78} & 64.51 & \textbf{0.69} & \textbf{0.66} & 0.64 & \textbf{0.68} & 0.73 & \textbf{0.72} & \textbf{69.89} & \textbf{0.74} & \textbf{0.79} & 0.76 & 0.57 & \textbf{0.75} & \textbf{0.83}\\ \hline
\end{tabular}}
\end{table*}
\myparagraph{Ablation study.}
In order to capture the gradual improvement of our framework through different stages, we conducted a serial ablation study and built two sub-variants of our frameworks. 1) Variant-1: This variant uses only multi-scale GRU cells which concatenate the inputs from two distinct patches - $Pp$, $Np$ to generate the fused representation. Both the Correlation module and the Attention module were removed from our framework. Though neighboring patches were taken into consideration, we do not exploit the explicit correlation between $Pp$, $Np$ and between $Pp$, $Rp$. Also, the encoder output vector does not consider the relative importance of hidden states generated for multiple timepoints. 2) Variant-2: This variant consists of the Correlation module. However, the Attention module is neglected and equal importance is assigned to all the zone patches collected from multiple timepoint' images.
Thus we gradually zeroed into our framework which outperforms the sub-variants by a large margin in most zones. The results in Tables \ref{Left results} and \ref{Right results} suggest that exploiting the correlation between the nearby zones and remote zone patches leads to an increase in prediction performance. Moreover, the use of an attention layer to provide individual patch importance further boosts the accuracy. As an example, it can seen that for the left lung middle zone, our $M_{L_1}$ accuracy is 72.04\% while for Variant-1 and Variant-2 it is 67.74\% and 70.96\%, respectively. 

\myparagraph{Testing with $d-2$ timepoints as encoder input.}
We designed an experimental setup to analyze how the framework performs when patches from only first $d-2$ images are used as input to our GRU encoder. However, the task is still to predict the severity scores of $I_{t_{d}}$ image. From Supplementary Table 1, we can observe that even in this experimental setting, our model achieves highest accuracies for $L_1$, $L_2$, $R_1$ and $R_3$ - 64.51, 62.36, 66.67 and 59.13, while achieving competitive scores for the other two zones. This suggests that our framework can perform well even if we have fewer number of timepoints as encoder input.
\section{Conclusion}
COVID-19 CXRs reveal varied spatial correlations among the lung infiltrates across different zones. Adjacent zones are generally found to be more correlated than two distant regions. We build a multi-scale GRU based encoder-decoder framework which accepts multiple inputs from different lung zones at a single timepoint. Unlike generative approaches, our model does not require registration between images from different timepoints. A novel two component correlation loss is introduced to explore the spatial correlations within nearby and distant lung fields in latent representation. Finally we use an attention layer to judge the relative importance of the images from available timepoints for computing the disease severity score at a future timepoint. 
\subsubsection{Acknowledgment:} Reported research was supported by the OVPR and IEDM seed grants, 2020 at Stony Brook
University, NIGMS T32GM008444, and NIH 75N92020D00021 (subcontract). The content is solely the responsibility of the authors and does not necessarily represent the official views of the National Institutes of Health.

\bibliographystyle{splncs04}
\bibliography{bibliography}




\end{document}


\title{Supplementary Material of \\ Attention-based Multi-scale Gated Recurrent Encoder with Novel Correlation Loss for COVID-19 Progression Prediction}
\titlerunning{Multi-scale GRU Encoder framework with Correlation Loss}

\author{Aishik Konwer\inst{1}, Joseph Bae\inst{2}, Gagandeep Singh\inst{3}, Rishabh Gattu\inst{3}, Syed Ali\inst{3}, Jeremy Green\inst{3}, Tej Phatak\inst{3}, Prateek Prasanna\inst{2}}


\authorrunning{A. Konwer et al.}

\institute{Department of Computer Science, Stony Brook University, NY, USA
\and Department of Biomedical Informatics, Stony Brook University, NY, USA
\and Department of Radiology, Newark Beth Israel Medical Center, NJ, USA
\email{akonwer@cs.stonybrook.edu, prateek.prasanna@stonybrook.edu}}

\maketitle{}  

\section{Additional Experiments}
%
\begin{table*}[ht]
\caption{Quantitative Results showing accuracies on lung zones using CXRs from $d-2$ timepoints as input to Encoder}
\label{Lung t-1}
\center
\resizebox{.7\textwidth}{!}{\begin{tabular}{|c|c|c|c|c|c|c|}
\hline
\multirow{2}{*}{\textbf{Methods}} & \multicolumn{3}{c|}{\textbf{Left lung $Acc (\%)$}} & \multicolumn{3}{c|}{\textbf{Right lung $Acc (\%)$}}  \\ \cline{2-7} 
 & Upper & Middle & Lower & Upper & Middle & Lower\\ \hline
Baseline-1 & 58.06 & 60.21 & 52.68 & 61.29 & 52.68 & 53.76 \\ \hline
Baseline-2 & 59.13 & 52.68 & 54.83 & 54.83 & 56.98 & 49.46 \\ \hline
\begin{tabular}[c]{@{}c@{}}Variant-1 \end{tabular} & 54.83 & 56.98 & 53.76 & 58.06 & 53.76 & 55.91\\ \hline
Variant-2 & \textbf{64.51} & 60.21 & \textbf{58.06} & 63.44 & \textbf{62.36} & 58.06\\ \hline
Our Approach & \textbf{64.51} & \textbf{62.36} & 56.98 & \textbf{66.67} & 61.29 & \textbf{59.13} \\ \hline
\end{tabular}}
\end{table*}

\section{Comparison of computational complexity}
We implemented our framework on a server with 11gb Nvidia RTX 2080 Ti gpu. Each model in the proposed approach was trained in ~3.4 hours for 30 epochs. Baseline 1 and 2 took ~2 hours and 1.25 hours respectively.

\section{Additional dataset details}
CXRs taken from Stony Brook University Hospital were acquired using the portable DRX Revolution machine developed by Carestream Health with AP image technique. CXRs taken from Newark Beth Israel Medical Center were acquired using GE Optima XR240 AMX portable machines. Each lung zone severity score was determined by agreement among three expert readers $(\geq15$, $\geq3$, and $\geq2$ years
of experience, respectively).

\section{Baseline Architecture}
\begin{figure}[h]

  	\begin{minipage}[b]{1.0\linewidth}
  		\centering
  		\centerline{\includegraphics[width= 4 in]{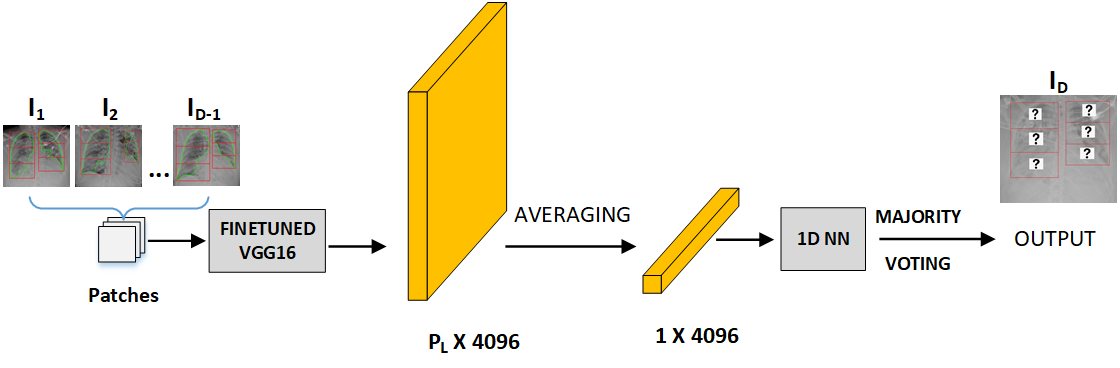}}
  	\end{minipage}

  	\caption{Architecture of the baseline approach
  }
  	\label{fig:baseline}
\end{figure}




